\documentclass{sig-alternate-05-2015}

\usepackage{times}
\usepackage{ctable}
\usepackage{multirow}
\usepackage{balance}
\usepackage{booktabs}
\usepackage{hyphsubst}
\usepackage{graphicx}
\usepackage{amsmath}
\usepackage{url}
\usepackage{breakurl}
\usepackage{hhline}
\usepackage[caption = false]{subfig}

\newcounter{lizcounter}
\DeclareRobustCommand{\liz}[1]{\textbf{/* #1 (liz) */}\stepcounter{lizcounter}\typeout{LaTeX Warning: liz comment \thelizcounter: #1 (line \the\inputlineno)}}
\DeclareMathOperator*{\argmax}{arg\,max}

\newcounter{findingscounter}
\newcommand{\findingbox}[1]{%
	\stepcounter{findingscounter}%
	\vspace{-1ex} %
	\vspace{2mm}
	\begin{center} %
		\noindent %
		\framebox{\parbox{.97\columnwidth}{%
			\textbf{Finding~\thefindingscounter:} #1}}%
	\end{center} %
	\vspace{-1ex}}

\newcommand{\para}[1]{\vspace{2mm}\noindent\textbf{#1}}
\newcommand{\subpara}[1]{\textit{\textbf{#1}}}

\begin{document}

\title{Temporal Effects on Hashtag Reuse in Twitter:\\A Cognitive-Inspired Hashtag Recommendation Approach}

\numberofauthors{3}
\author{
\alignauthor
Dominik Kowald\\
			 \affaddr{Know-Center}\\
       \affaddr{Graz, Austria}\\
       \email{dkowald@know-center.at}		
\alignauthor
Subhash Pujari\\
			 \affaddr{Graz University of Technology}\\
       \affaddr{Graz, Austria}\\
       \email{s.pujari@student.tugraz.at}
\alignauthor
Elisabeth Lex\\
			 \affaddr{Graz University of Technology}\\
       \affaddr{Graz, Austria}\\
       \email{elisabeth.lex@tugraz.at}
}

\conferenceinfo{}{Copyright is held by the authors.}

\maketitle

\begin{abstract}
Hashtags have become a powerful tool in social platforms such as Twitter to categorize and search for content, and to spread short messages across members of the social network. In this paper, we study temporal hashtag usage practices in Twitter with the aim of designing a cognitive-inspired hashtag recommendation algorithm we call BLL$_{I,S}${}. Our main idea is to incorporate the effect of time on (i) individual hashtag reuse (i.e., reusing own hashtags), and (ii) social hashtag reuse (i.e., reusing hashtags, which has been previously used by a followee) into a predictive model. For this, we turn to the Base-Level Learning (BLL) equation from the cognitive architecture ACT-R, which accounts for the time-dependent decay of item exposure in human memory. We validate BLL$_{I,S}${} using two crawled Twitter datasets in two evaluation scenarios: firstly, only temporal usage patterns of past hashtag assignments are utilized and secondly, these patterns are combined with a content-based analysis of the current tweet. In both scenarios, we find not only that temporal effects play an important role for both individual and social hashtag reuse but also that BLL$_{I,S}${} provides significantly better prediction accuracy and ranking results than current state-of-the-art hashtag recommendation methods.
\end{abstract}

\para{Keywords.} Twitter; Hashtags; BLL equation; ACT-R; TF-IDF; Recency; Hashtag Recommendation; Hashtag Reuse Prediction

\section{Introduction} \label{sec:intro}
Over the past years, the microblogging platform Twitter has become one of the most popular social networks on the Web. Users can build a network of follower connections to other Twitter users, which means that they can subscribe to content posted by their \textit{followees} \cite{Myers2014,kwak2010}. Twitter was also the first social platform that adopted the concept of \textit{hashtags}, as suggested by Chris Messina\footnote{\url{https://twitter.com/chrismessina/status/223115412}}.

Hashtags are freely-chosen keywords starting with the hash character ``\#'' to annotate, categorize and contextualize Twitter posts (i.e., tweets) \cite{romero2011,Huang2010}. The advantage of hashtags is that anyone with an interest in a hashtag can track it and search for it \cite{small2011hashtag}, thus receiving content posted by somebody outside of their own Twitter network. For example, users can retrieve tweets created during the European football championship by searching for the hashtag \textit{\#euro2016}, even if they do not have a social link to the tweet producers. Meanwhile, many social platforms, such as Instagram and Facebook, have adopted hashtags as well.

\para{Problem.} Unsurprisingly, the widespread acceptance of hashtags has sparked a lot of research in the field of \textit{hashtag recommendations} (see Section \ref{sec:relatedwork} for a selection of approaches) to support users in assigning the most descriptive hashtags to their posts. Existing methods typically utilize collaborative, content and topic features of tweets to recommend hashtags to users. Undoubtedly, these features play an important role in recommending hashtags that best describe a tweet. In this paper, however, we are especially interested in predicting which hashtags a user will likely apply in a newly created tweet given previous hashtag assignments.

The main problem we want to address is whether we can identify \textit{temporal usage patterns} that influence if a Twitter user will likely utilize a certain hashtag in a tweet, given the hashtags she and/or her followees have been using in the past. Our goal is to describe such temporal usage patterns using a model from human memory theory and to design a hashtag recommendation algorithm based on that. To the best of our knowledge, so far, few studies (e.g., \cite{harvey2015long}) have investigated the way temporal effects can be exploited in the hashtag recommendation process.

\para{Approach and methods.} We propose a cognitive-inspired hashtag recommendation algorithm we call BLL$_{I,S}${} that is based on temporal usage patterns of hashtags derived from empirical evidence. In essence, these patterns reflect how a person's own hashtags as well as hashtags from the social network are utilized and reused. In our approach, we utilize the Base-Level Learning (BLL) equation from the cognitive architecture ACT-R \cite{anderson2004integrated,anderson_reflections_1991} to model temporal usage of hashtags. The BLL equation accounts for the time-dependent decay of item exposure in human memory. It quantifies the usefulness of a piece of information (e.g., a hashtag) based on how frequently and how recently it was used by a user in the past and models this time-dependent decay by means of a power-law distribution. Thus, BLL$_{I,S}${} takes into consideration the frequency and recency of hashtags used by a user and her followees in the past. 

We presented the BLL equation in our previous work as a model to recommend tags in social bookmarking systems such as BibSonomy and CiteULike \cite{www_bll, Kowald2016a}. In the present work, we build upon these results by adopting the BLL equation to model the effect of time on the reuse of individual and social hashtags to build our hashtag recommendation algorithm. We demonstrate the efficacy of our approach in two empirical social networks crawled from Twitter. The first social network, termed \textit{CompSci}{} dataset, is built upon the tweets of a sample of Twitter users, who have been identified as computer scientists in previous related work \cite{hadgu2014identifying}, and their followees. The second network, termed \textit{Random}{} dataset, is built upon the tweets of a set of randomly chosen Twitter users and their followees. We experiment with these datasets to investigate the performance of our hashtag recommendation approach in two settings: (i) tweets of a domain-specific Twitter network, and (ii) tweets of a random network of Twitter users.

\para{Contributions and findings.} The main contributions of our work are two-fold. Firstly, our paper shows that time has a large effect on individual as well as social hashtag reuse in Twitter. Specifically, we observe a time-dependent decay of individual and social hashtag reuse that follows a power-law distribution. This finding paves the way for our idea to utilize the BLL equation as a predictive model to recommend hashtags for new tweets. Thus, our second contribution is that we design, develop and evaluate a personalized hashtag recommendation algorithm based on the BLL equation that outperforms current state-of-the-art approaches.

We implement the BLL equation in two variants, where the first one (i.e., BLL$_{I,S}${}) predicts the hashtags of a user solely based on past hashtag usage, and the second one (i.e., BLL$_{I,S,C}${}) combines BLL$_{I,S}${} with a content-based tweet analysis to also incorporate the text of the currently proposed tweet of a user. We evaluate our approach using standard evaluation protocols and metrics, and we find that our approach provides significantly higher prediction accuracy and ranking estimates than current state-of-the-art hashtag recommendation algorithms in both scenarios. We attribute this to the fact that our approach, in contrast to other related methods, mimics the way humans use and adapt hashtags by building upon insights from human memory theory (i.e., the BLL equation).

\para{Structure of this paper.} In Section \ref{sec:datasets}, we continue by describing the crawling procedure of our two Twitter datasets and analyzing hashtag usage types in these datasets. Then, in Section \ref{sec:analysis}, we study temporal usage patterns of individual and social hashtag reuse. In Section \ref{sec:approach}, we describe two variants of our approach (i.e., without and with the current tweet). This is followed in Section \ref{sec:evaluation} by our evaluation methodology and experimental results. Finally, we discuss related work in the field in Section \ref{sec:relatedwork} and we give a summary of our findings as well as our future plans in Section \ref{sec:conclusion}.

\section{Datasets} \label{sec:datasets}
In this section, we describe the data collection procedure and the two datasets we use for our study. Additionally, we investigate individual as well as social hashtag reuse patterns in our datasets as a prerequisite for our hashtag recommendation approach.

\para{Crawling strategy and dataset statistics.} In order to address our research goals, we crawl two datasets using the Search API of Twitter\footnote{\url{https://dev.twitter.com/rest/public/search}}. The final statistics of these datasets are illustrated in Table \ref{tab:datasets}.

The first one (i.e., \textit{CompSci}{} dataset) consists of researchers from the field of computer science and their followees, while the second one (i.e., \textit{Random}{} dataset) consists of random people and their followees. Our idea is to test our hashtag recommendation approach in two different network settings: (i) a domain-specific one, in our case the domain of computer scientists, and (ii) a more general one consisting of random Twitter users. Our crawling strategy for both datasets comprises of the following four steps:

\subpara{(a) Crawl seed users.} We start with identifying and crawling a list of seed users $U_S$ for each dataset. In the case of the \textit{CompSci}{} dataset, we take the users who were identified as computer scientists in the work of \cite{hadgu2014identifying}. In the case of the \textit{Random}{} dataset, we used the Streaming API of Twitter\footnote{\url{https://dev.twitter.com/streaming/overview}} in October 2015 to get a stream of tweets and extracted the user-ids to get our list of random seed users. From both user lists, we remove all users with more than 180 followees, which results in $|U_S|$ = 2,551 seed users for the \textit{CompSci}{} dataset and $|U_S|$ = 3,466 seed users for the \textit{Random}{} dataset. The threshold of using a maximum of 180 followees is chosen because the Twitter Search API only allows 180 requests per 15 minutes, which gives us the possibility to crawl the tweets of all followees of a seed user within this reasonable time window.

\subpara{(b) Crawl followees.} Next, we use these follower relationships to crawl the followees $F$ of the seed users in order to create a directed user network for analyzing the social influence on hashtag reuse. Based on the number of seed users, the average number of followees per seed user $|F| / |U_S|$ = 94 in the case of the \textit{CompSci}{} dataset and 72 in the case of the \textit{Random}{} dataset. Following these notations, the set of followees of user $u$ is denoted as F$_u$ in the remainder of this paper. Overall, our crawling procedure gives us $|U|$ = 91,776 total users for the \textit{CompSci}{} dataset and $|U|$ = 127,112 total users for the \textit{Random}{} dataset.

\begin{table}[t!]
	\small
  \setlength{\tabcolsep}{2.3pt}	
  \centering
    \begin{tabular}{l||cccccc}
    \specialrule{.2em}{.1em}{.1em}
											Dataset				& $|U_S|$					& $|F|$				& $|U|$				& $|T|$					& $|HT|$			& $|HTAS|$				\\\hline 
											\textit{CompSci}{}				&	2,551						&	241,225			&	91,776			&	5,649,359			&	1,081,403		&	9,161,842					\\\hline
											\textit{Random}{}				& 3,466		  			&	252,219			& 127,112 		& 8,157,702 		& 1,507,773		& 13,628,750				\\
		\specialrule{.2em}{.1em}{.1em}								
    \end{tabular}
    \caption{Statistics of our \textit{CompSci}{} and \textit{Random}{} Twitter datasets. Here, $|U_S|$ is the number of seed users, $|F|$ is the number of followees of these seed users, $|U|$ is the number of total users, $|T|$ is the number of Tweets, $|HT|$ is the number of distinct hashtags and $|HTAS|$ is the number of hashtag assignments.}
  \label{tab:datasets}
\end{table}

\begin{figure}[t!]
   \centering
      \includegraphics[width=0.48\textwidth]{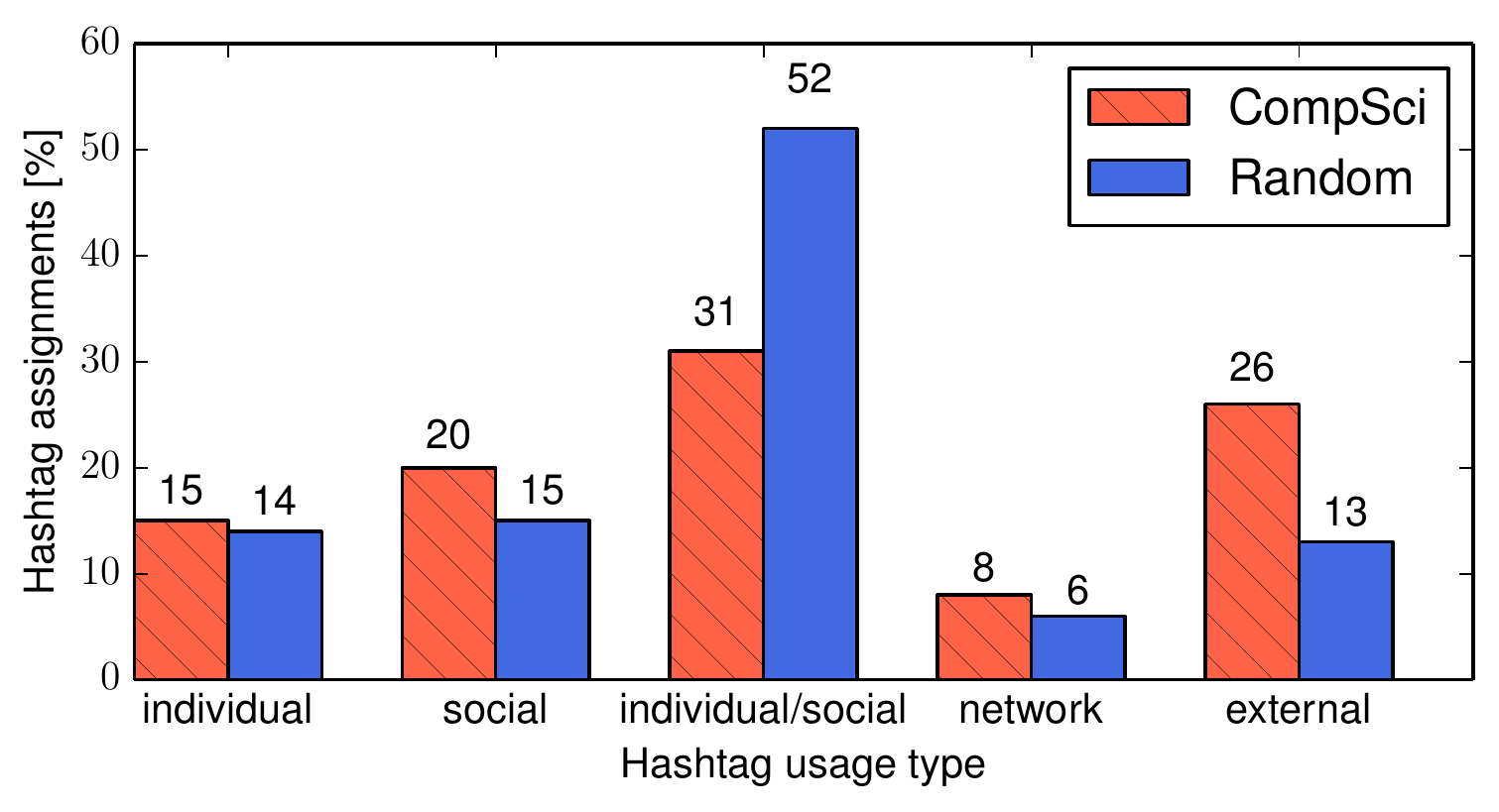} 
   \caption{Analysis of hashtag usage types in our two datasets. For each hashtag assignment, we study whether the corresponding hashtag has been used by the same user before in time (``individual''), by some of the users she follows (``social''), by both (``individual/social''), by anyone else in the dataset (``network'') or neither of them (``external''). We find that between 66\% and 81\% of hashtag assignments in both datasets can be explained by individual or social hashtag usage (i.e., the sum of ``individual'', ``social'' and ``individual/social'').
\vspace{-3mm}}
	 \label{fig:intro}
\end{figure}

\begin{figure*}[t!]
   \centering
	 \captionsetup[subfigure]{justification=centering}
	 \subfloat[][Individual hashtag reuse\\\textit{CompSci}{} dataset ($R^2$ = .883)]{ 
      \includegraphics[width=0.24\textwidth]{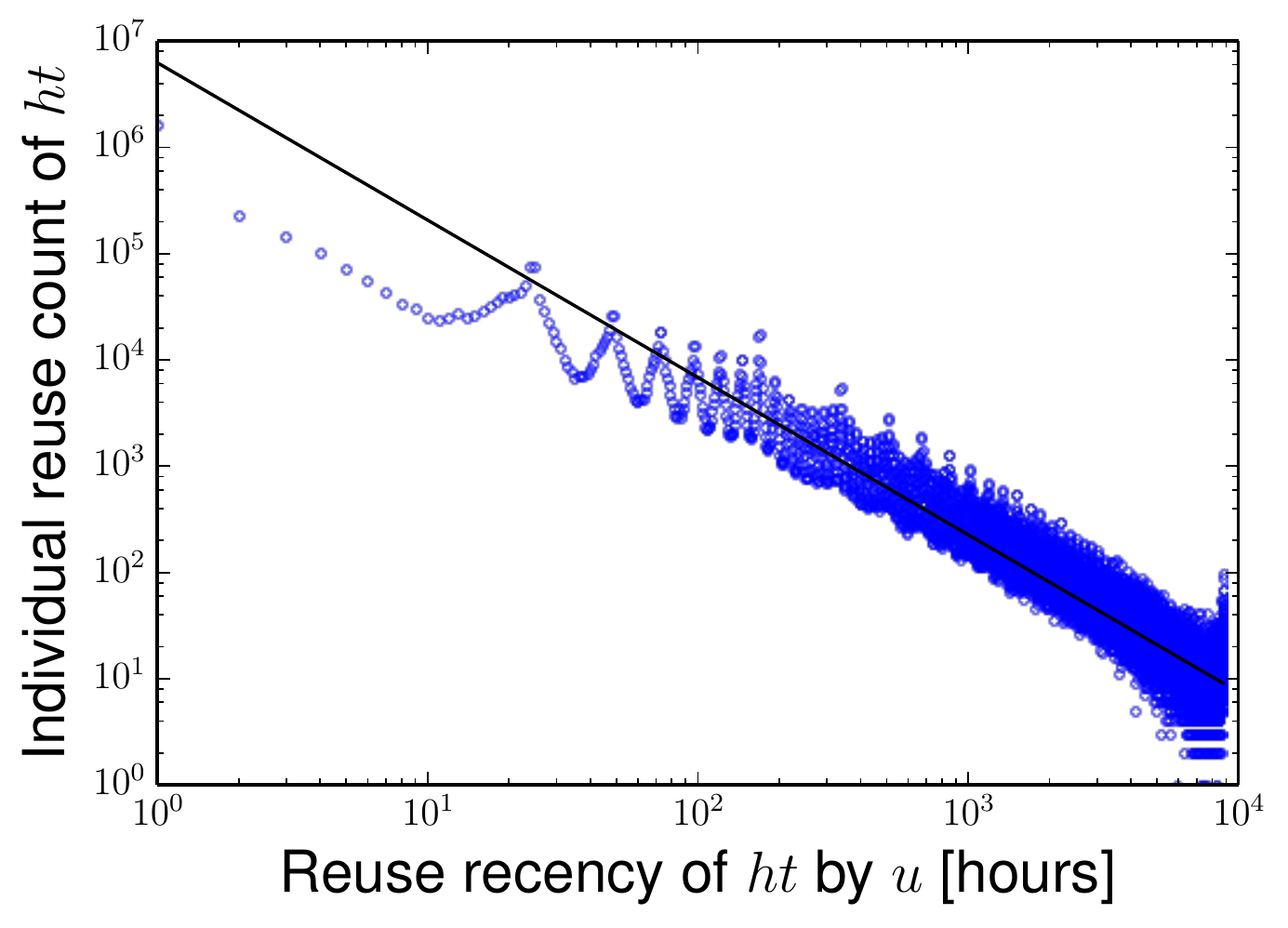} 
   }
	 \subfloat[][Individual hashtag reuse\\\textit{Random}{} dataset ($R^2$ = .894)]{ 
      \includegraphics[width=0.24\textwidth]{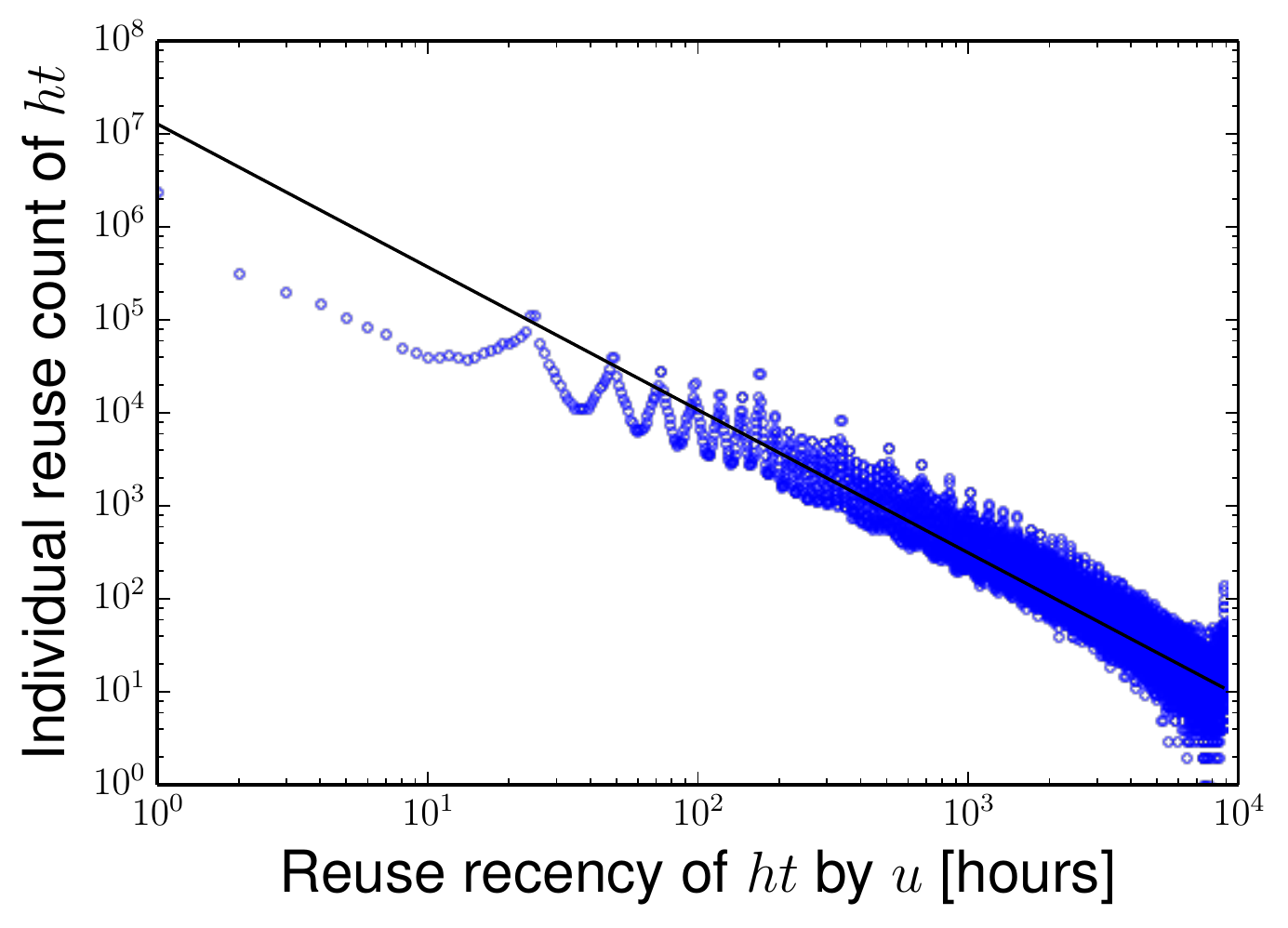} 
   }
	 \subfloat[][Social hashtag reuse\\\textit{CompSci}{} dataset ($R^2$ = .689)]{ 
      \includegraphics[width=0.24\textwidth]{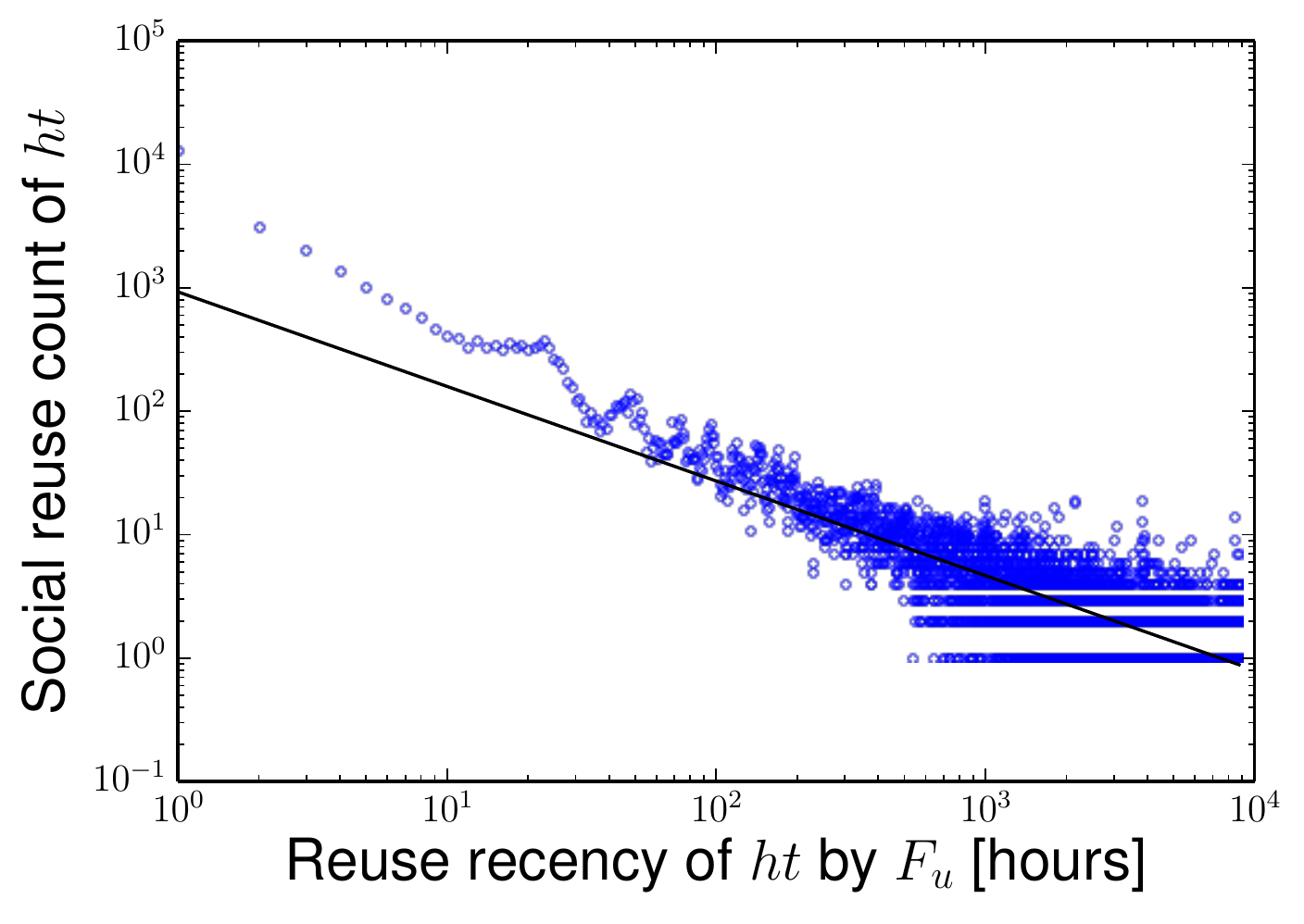} 
   }  
	 \subfloat[][Social hashtag reuse\\\textit{Random}{} dataset ($R^2$ = .771)]{ 
      \includegraphics[width=0.24\textwidth]{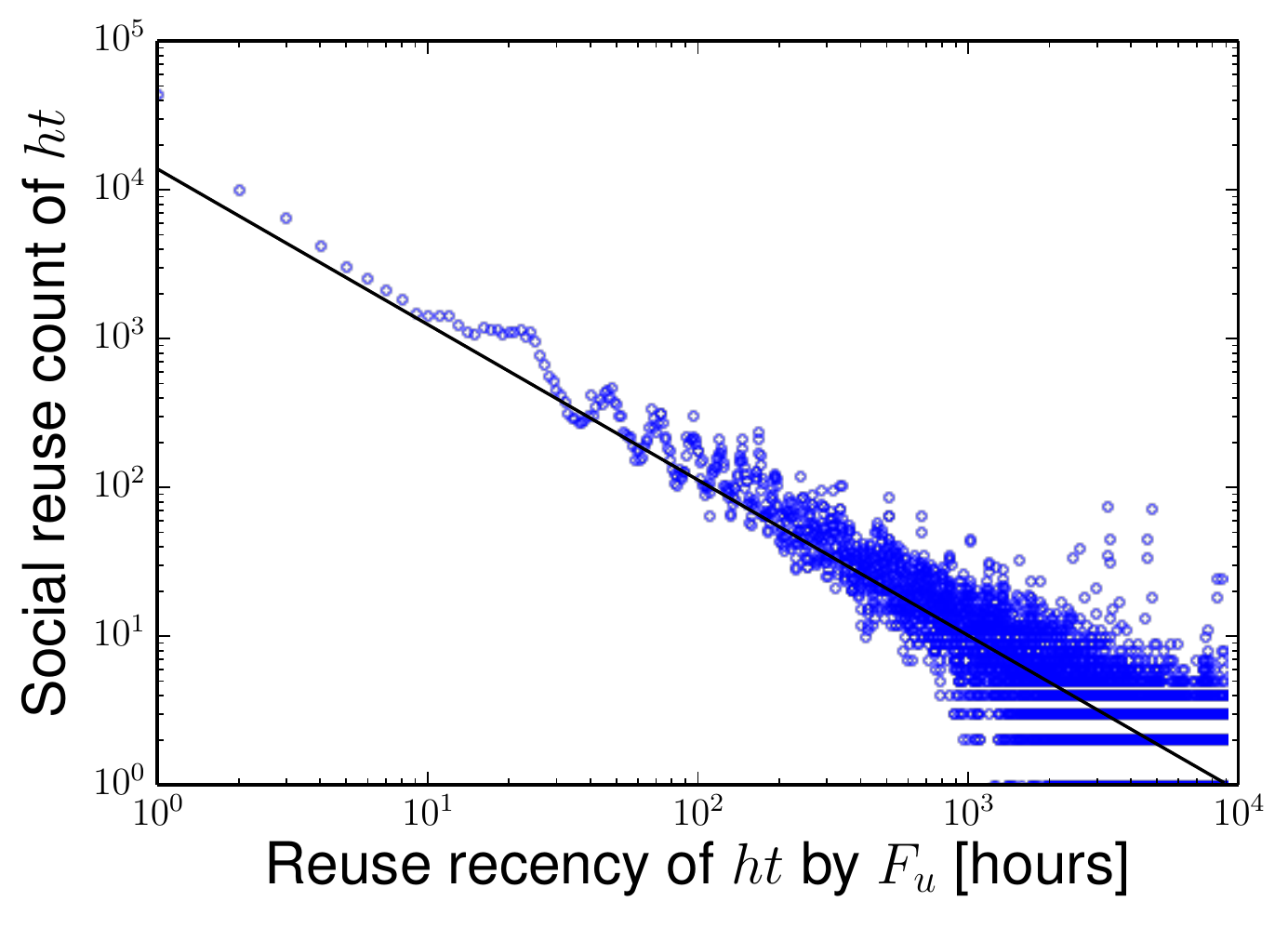} 
   }
   \caption{The effect of time on individual and social hashtag reuse for the \textit{CompSci}{} and \textit{Random}{} datasets (plots are in log-log scale). Plots (a) and (b) show that the more recently a hashtag $ht$ was used by a user $u$, the higher its individual reuse count (i.e., people tend to reuse hashtags that have been used very recently by their own). Plots (c) and (d) show that the more recently a user $u$ was exposed to a hashtag $ht$, which was used by her followees $F_u$, the higher its social reuse count (i.e., people tend to reuse hashtags that have been used recently in the social network). Additionally, we report the $R^2$ estimates for the linear fits of the data. We find that temporal effects play an important role in individual and social hashtag reuse in both datasets.
\vspace{-3mm}}
	 \label{fig:analysis}
\end{figure*}

\subpara{(c) Crawl tweets.} In the third step, we crawl the 200 most recent tweets of all the users and remove the tweets in which no hashtags are used. The threshold of a maximum of 200 most recent tweets is set because of another restriction of the Twitter Search API that only allows 200 tweets to be received per a single request. This crawling procedure results in $|T|$ = 5,649,359 tweets for the \textit{CompSci}{} dataset with an average number of tweets per user $|T| / |U|$ = 61, and $|T|$ = 8,157,702 tweets for the \textit{Random}{} dataset with $|T| / |U|$ = 64. Our crawled tweets cover a time range from 2007 to 2015.

\subpara{(d) Extract hashtags.} Finally, we extract the hashtags of the tweets by searching for all words that start with a ``\#'' character. This results in $|HTAS|$ = 9,161,842 hashtag assignments for $|HT|$ = 1,081,403 distinct hashtags in the \textit{CompSci}{} network and $|HTAS|$ = 13,628,750  for $|HT|$ = 1,507,773 in the \textit{Random}{} network. Thus, in both datasets, each distinct hashtag is used approximately 9 times on average and each user uses approximately 100 hashtag assignments in her tweets on average. Examples for popular hashtags are \textit{\#bigdata}, \textit{\#iot} and \textit{\#ux} in case of the \textit{CompSci}{} dataset, and \textit{\#shahbag}, \textit{\#ff} and \textit{\#art} in case of the \textit{Random}{} dataset.

\para{Analysis of hashtag usage types.} In our datasets, we analyze hashtag assignments as well as hashtag reuse practices with the aim of identifying the different types of hashtag usages as a prerequisite for our recommendation approach. Specifically, for each hashtag assignment, we study whether the corresponding hashtag has either been used by the same user before (``individual''), by some of her followees (``social''), by both (``individual/social''), by anyone else in the dataset (``network'') or by neither of them (``external'').

The results of this study are shown in Figure \ref{fig:intro}. We find that 66\% of hashtag assignments in the \textit{CompSci}{} dataset and 81\% in the \textit{Random}{} dataset can be explained by individual or social hashtag reuse. This finding further corroborates our choice to utilize these two types of influences (i.e., individual and social) to create our model. In contrast to these large numbers, the 6\% to 8\% of hashtags in the ``network'' category is relatively small. Interestingly, the amount of ``external'' hashtags is twice as high in the \textit{CompSci}{} dataset (i.e., 26\%) as in the \textit{Random}{} one (i.e., 13\%). Thus, in our datasets, computer scientists tend to use more hashtags, which have not been previously introduced in the network, than random Twitter users. Because of this, we believe that the recommendation accuracy results would generally be lower in the \textit{CompSci}{} dataset than in the \textit{Random}{} one, which will be evaluated in Section \ref{sec:evaluation}. Summing up, both individual and social hashtags have an impact on users' choice of hashtags for a new tweet.

\section{Temporal Effects on Hashtag\\Reuse in Twitter} \label{sec:analysis}
In this section, we study to what extent temporal effects play a role in the reuse of individual and social hashtags in our two datasets (i.e., \textit{CompSci}{} and \textit{Random}{}). Specifically, we analyze the recency of hashtags assignments (i.e., the time since the last hashtag usage/exposure), as well as whether this effect of time-dependent decay follows a power-law or exponential distribution.

\para{Temporal effects on individual hashtag reuse.} The effect of time on individual hashtag reuse is visualized in the plots (a) and (b) of Figure \ref{fig:analysis}. To put the x-scale of these plots onto a meaningful range, we set the threshold for the maximum hashtag reuse recency to one year (i.e., 8,760 hours). The plots show the individual hashtag reuse count plotted over the reuse recency of a hashtag $ht$ by a user $u$ in hours. Hence, for each hashtag assignment of a hashtag $ht$ by user $u$, we take the time since the last usage of $ht$ by $u$ (i.e., the reuse recency) and pool together all hashtag assignments with the same recency value (i.e., the same time difference in hours). The individual reuse count for this recency value is then given by the size of the set of these hashtag assignments.

The two plots show similar results for both datasets and indicate that the more recently a hashtag $ht$ was used by a user $u$ in the past, the higher its individual reuse count is. Interestingly, there is a clear peak after 24 hours in both datasets, which further indicates that users typically use the same set of hashtags in this time span and thus, tend to tweet about similar topics on a daily basis. Furthermore, we also observe high $R^2$ values of nearly .9 for the linear fits in the log-log scaled plots, which indicates that a large amount of our data can be explained by a power function. This is also suggested by the power-law-based model of the BLL equation \cite{anderson_reflections_1991,anderson2004integrated}. In contrast, the linear fits in log-linear scaled plots only provide $R^2$ values of approximately .7, where high values would speak in favor of an exponential function.

\para{Temporal effects on social hashtag reuse.} Plots (c) and (d) of Figure \ref{fig:analysis} show the effect of time on the social hashtag reuse for the \textit{CompSci}{} and \textit{Random}{} datasets. These plots are created similarly as plots (a) and (b) but this time, we plot the social hashtag reuse count over the reuse recency of a hashtag $ht$ by the followees $F_u$ of user $u$. Hence, for each hashtag assignment of $ht$ by $u$, we take the most recent usage timestamp of $ht$ by $F_u$. The difference between this timestamp and the timestamp of the currently analyzed hashtag assignment indicates the time since the last social exposure of $ht$ to $u$. Again, we set the threshold for the maximum hashtag reuse recency to one year (i.e., 8,760 hours).

In these plots, we observe similar results for the two datasets since, in both cases, the more recently a user was exposed to a hashtag, the higher its social reuse count is. Furthermore, there is again (i) a clear peak after 24 hours, and (ii) the $R^2$ values for the linear fits in the log-log scaled plots (i.e., = .7) are larger than in the log-linear scaled plots (i.e., = .4), which speaks in favor of a power function. We now study if this is really the case.

\para{Power-law vs. exponential time-dependent decay.} The question whether a power or an exponential function is better suited to model the time-dependent decay of hashtag reuse is of interest especially for the design of our hashtag recommendation approach since both types of functions have been used in the area of time-aware recommender systems. While the BLL equation suggests the use of a power function to model the decay of item exposure in human memory \cite{anderson_reflections_1991}, related hashtag recommender approaches, such as the one proposed in \cite{harvey2015long}, use an exponential function for this purpose. As already mentioned, the visual inspection of Figure \ref{fig:analysis} and the $R^2$ values of the linear fits favor a power function. However, \cite{clauset2009power} has shown that this least squares-based method can lead to misinterpretations and thus, a likelihood ratio-based test is suggested.
 
We use the Python implementation \cite{alstott2014powerlaw} of the method described in \cite{clauset2009power} to validate if a power function produces a better fit than an exponential one. The results of this test are shown in Table \ref{tab:analysis}. The main value of interest here is the log-likelihood ratio $R$ between the two functions. As we see, $R > 0$ in all four cases with $p < .001$. This means that the power function indeed provides a better fit than the exponential function for explaining temporal effects on individual and social hashtag reuse. We also provide the $x_{min}$ and $\alpha$ values of the fits. In this respect, the $\alpha$ slopes can be used to set the $d$ parameter of the BLL equation (i.e., 1.7 in the individual case and 1.25 in the social case, see Section \ref{sec:approach}). Interestingly, these values are much higher than the suggested value of BLL's $d$ parameter, which is .5 \cite{anderson2004integrated}. We believe that this is the case because tweeting is more strongly influenced by temporal interest drifts than other applications studied in the ACT-R community (e.g., \cite{anderson_reflections_1991}).

\vspace{-1mm}
\findingbox{Temporal effects have an important influence on both individual as well as social hashtag reuse: people tend to reuse hashtags that were used very recently by their own and/or by their Twitter followees. Furthermore, a power function is better suited to model this time-dependent decay than an exponential one. This suggests that the BLL equation from the cognitive architecture ACT-R should be a suitable model for designing our time-dependent hashtag recommendation algorithm.}

\begin{table}[t!]
	\small
  \setlength{\tabcolsep}{7.2pt}	
  \centering
    \begin{tabular}{l||l|cc}
    \specialrule{.2em}{.1em}{.1em}
											Dataset				& Parameter								& Individual ht reuse 	& Social ht reuse	\\\hline 
											\multirow{3}{*}{\centering{\textit{CompSci}{}}}						 
																		&	x$_{min}$								& 141				 						& 1							\\
																		& $\alpha$								& 1.699			 						& 1.242					\\
																		& $R$											&	\textbf{188}					& \textbf{164}	\\\hline																		
											\multirow{3}{*}{\centering{\textit{Random}{}}}														
																		&	x$_{min}$								& 141				 						& 1							\\
																		& $\alpha$								& 1.723			 						& 1.269					\\
																		& $R$											&	\textbf{235}					& \textbf{294}	\\			
		\specialrule{.2em}{.1em}{.1em}								
    \end{tabular}
    \caption{Power-law vs. exponential time-dependent decay. We see that a power function provides a better fit than an exponential function ($R > 0$) for explaining temporal effects on individual and social hashtag reuse in our two datasets ($p < .001$).
\vspace{-5mm}}	
  \label{tab:analysis}
\end{table}\section{A Cognitive-Inspired Hashtag\\Recommendation Approach} \label{sec:approach}
In the previous section, we have shown that temporal effects are important factors when users reuse individual and social hashtags. In this section, we use these insights as a basis to design our hashtag recommendation approach illustrated in Figure \ref{fig:approach}. Thus, we distinguish between hashtag recommendations without (\textit{Scenario 1}{}) and with (\textit{Scenario 2}{}) incorporating the current tweet $t$.

Whereas the first variant of our approach solely uses the past hashtags of a user $u$ and/or her followees F$_u$, the second variant also utilizes the text of the current tweet $t$. Hence, these two scenarios also differ in their possible use cases since the first one aims to foresee the topics a specific user will tweet about based on the predicted hashtags, whereas the second one aims to support a user in finding the most descriptive hashtags for a new tweet text \cite{Godin2013}.

For reasons of reproducibility, we implement and evaluate our approach by extending our open-source tag recommender benchmarking framework \textit{TagRec}. The source code and framework is freely accessible for scientific purposes on the Web\footnote{\url{https://github.com/learning-layers/TagRec}}.

\subsection{Scenario 1: Hashtag rec. w/o current tweet}
\begin{figure}[t!]
   \centering
      \includegraphics[width=0.48\textwidth]{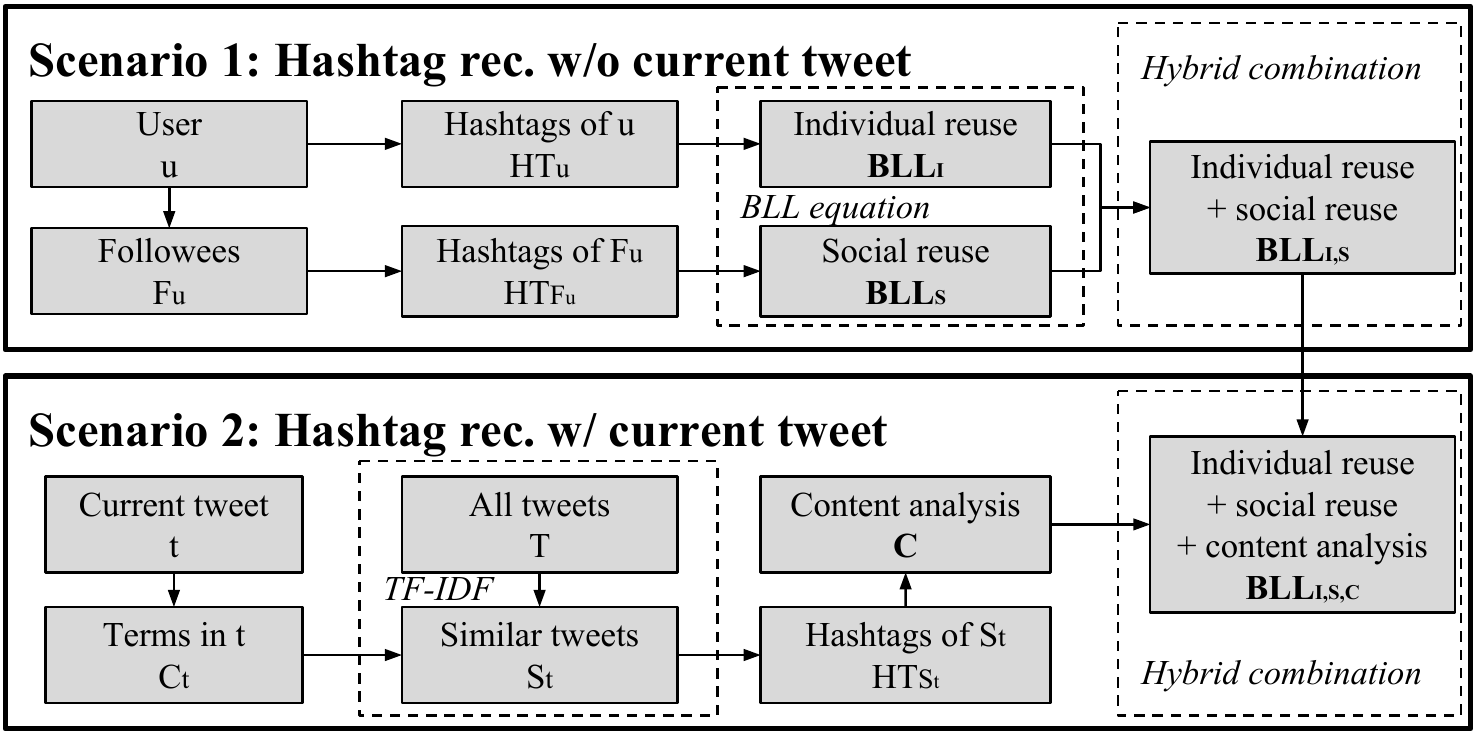} 
   \caption{Schematic illustration of our cognitive-inspired approach for hashtag recommendations. We implement our approach in two scenarios (i.e., without and with incorporating the content of the current tweet). In \textit{Scenario 1}{}, we use the BLL equation to realize (i) the individual BLL$_I$ algorithm, (ii) the social BLL$_S$ algorithm, and (iii) the hybrid BLL$_{I,S}${} algorithm, which combines both. In \textit{Scenario 2}{}, we use TF-IDF to identify similar tweets for a currently proposed tweet $t$ and identify the hashtags of the most similar ones. We combine this content-based tweet analysis with our BLL$_{I,S}${} method to provide personalized and content-aware hashtag recommendations in the form of our hybrid BLL$_{I,S,C}${} approach.
\vspace{-3mm}}
	 \label{fig:approach}
\end{figure}

For the first variant of our approach, we ignore the content of the current tweet $t$ and solely utilize past hashtag usages. As already stated, we use the BLL equation coming from the cognitive architecture ACT-R \cite{anderson2004integrated,anderson_reflections_1991} for this task. We go for a cognitive-inspired approach, since we know from research on the underlying mechanisms of social tagging that the way users choose tags for annotating resources (e.g., Web links) strongly corresponds to processes in human memory and its cognitive structures \cite{cress2013collective,seitlinger2015verbatim}. The BLL equation quantifies the general usefulness of a piece of information (e.g., a word or hashtag) by considering how frequently and recently it was used by a user in the past. Formally, it is given by:
\begin{equation}\label{eq:bll}
    B_i = ln(\sum\limits_{j = 1}\limits^{n}{t_{j}^{-d}})
\end{equation}
where $B_i$ is the base-level activation of a memory unit $i$ and $n$ is the frequency of $i$'s occurrences in the past (i.e., how often $i$ was used by $u$). Furthermore, $t_j$ states the recency (i.e., the time since the $j$th occurrence of $i$) and the exponent $d$ accounts for the power-law of time-dependent decay. As visualized in \textit{Scenario 1}{} of Figure \ref{fig:approach}, we adopt the BLL equation for (i) modeling the reuse of individual hashtags (BLL$_{I}$), (ii) modeling the reuse of social hashtags (BLL$_{S}$), and (iii) combining the former two into a hybrid recommendation approach (BLL$_{I,S}$).

\vspace{2mm} \noindent \textbf{Modeling individual hashtag reuse.}
In order to model the reuse of individual hashtags, we define the individual base-level activation $B_I(ht, u)$ of a hashtag $ht$ for a user $u$ as follows:
\begin{equation}
	B_I(ht, u) = \ln(\sum\limits_{j = 1}\limits^{n}{(TS_{ref} - TS_{ht,u,j})^{-d_I})}
\end{equation}
where $n$ denotes the number of times $ht$ was used by $u$ in the past (i.e., $|HTAS_{ht,u}|$) and the term $TS_{ref} - TS_{ht,u,j}$ states the recency of the $j$th usage of $ht$ by $u$. In this respect, $TS_{ref}$ is the reference timestamp (i.e., when recommendations should be calculated) and $TS_{ht,u,j}$ is the timestamp when $ht$ was used by $u$ for the $j$th time. Based on the results of our analysis presented in Table \ref{tab:analysis}, we set the individual time-dependent decay factor $d_I$ to 1.7.

\vspace{2mm} \noindent \textbf{Modeling social hashtag reuse.}
We model the reuse of social hashtags in a similar way but instead of analyzing how frequently and recently a hashtag $ht$ was used by user $u$, we analyze how frequently and recently $ht$ was used by the set of followees $F_u$ of $u$. Thus, we formulate the social base-level activation $B_S(ht, u)$ of $ht$ for $u$ as follows:
\begin{equation}
	B_S(ht, u) = \ln(\sum\limits_{j = 1}\limits^{m}{(TS_{ref} - TS_{ht,F_u,j})^{-d_S})}
\end{equation}
where $m$ is the number of times $ht$ was used by $F_u$ before the reference timestamp $TS_{ref}$ (i.e., $|HTAS_{ht,F_u}|$). The term $TS_{ref} - TS_{ht,F_u,j}$ states the recency of the $j$th exposure of $ht$ to $u$ caused by $F_u$, where $TS_{ht,F_u,j}$ is the timestamp when $ht$ was used by $F_u$ for the $j$th time. As when modeling the individual hashtag reuse, we set the social time-dependent decay factor $d_S$ based on the results of our analysis in Table \ref{tab:analysis} (i.e., to 1.25).

\para{Combining individual and social hashtag reuse.}
As we have formalized the individual as well as social hashtag reuse, we want to mix both components in form of a hybrid approach using a linear combination \cite{jaschke2008tag}. Hence, in order to be able to add the individual and social base-level activations $B_I(ht, u)$ and $B_S(ht, u)$, we have to map these values onto a common range of 0 to 1 that add up to 1. Therefore, we define the softmax functions $\sigma(B_I(ht,u))$ and $\sigma(B_S(ht,u))$ as proposed by \cite{mcauley2013hidden,www_bll}. This is given by:
\begin{equation} \label{eq:sm}
	\sigma(B_I(ht,u)) = \frac{\exp(B_I(t, u))}{\sum\limits_{ht' \in HT_{u}}{\exp(B_I(ht', u))}}
 \end{equation}
where $HT_u$ is the set of distinct hashtags used by $u$. For $B_S(ht, u)$, the softmax function $\sigma(B_S(ht,u))$ can be calculated in the same way but on the basis of $HT_{F_u}$ (i.e., the set of hashtags used by $u$'s followees $F_u$). 
Taken together, the combined base-level activation $B_{I,S}$ for our BLL$_{I,S}$ approach is given by:
\begin{equation}  \label{eq:hybrid}
	B_{I,S}(ht,u) = \beta \underbrace{\sigma(B_I(ht, u))}_{BLL_I} + (1 - \beta) \underbrace{\sigma(B_S(ht, u))}_{BLL_S}
\end{equation}
where the $\beta$ parameter can be used to give weights to the two components. Based on experimentation, we set $\beta$ to .5 to equally weigh the individual and social influence. As indicated in Equation \ref{eq:hybrid} and Figure \ref{fig:approach}, we can also calculate predictions either solely based on the individual hashtag reuse, referred as BLL$_I$, or the social hashtag reuse, referred as BLL$_S$.

\subsection{Scenario 2: Hashtag rec. w/ current tweet} \label{sec:hashtagrec}
As shown in \textit{Scenario 2}{} of Figure \ref{fig:approach}, the second variant of our approach aims to provide hashtag suggestions while also incorporating the content of the currently proposed tweet $t$. Thus, we build on the unpersonalized method proposed by \cite{zangerle2011recommending} to find hashtags of similar tweets and combine this method with our BLL$_{I,S}${} approach to generate personalized and content-aware recommendations.

\para{Content-based tweet analysis.} We analyze the content of tweets in order to find similar tweets for a target tweet $t$ and to extract the hashtags of these similar ones. Therefore, we incorporate the term frequency-inverse document frequency (TF-IDF) statistic, which identifies the importance of a term for a document in a collection of documents. TF-IDF can be further used to calculate the similarity between two documents $d$ and $\overline{d}$ by summing up the TF-IDF statistics of $d$'s terms in $\overline{d}$. When applying this statistic to Twitter, we treat tweets as documents and calculate the similarity between the target tweet $t$ and a candidate tweet $\overline{t}$ as follows:
\begin{equation}
    sim(t, \overline{t}) = \sum\limits_{c \in C_t}{n_{c,\overline{t}} \times \log(\frac{|T|}{|\{t': c \in t'\}|})}
\end{equation}
where $C_t$ are the terms in the text of target tweet $t$, $n_{c, \overline{t}}$ is the number of times $c \in C_t$ occurs in the candidate tweet $\overline{t}$, $|T|$ is the number of tweets in the dataset and $|\{t': c \in t'\}|$ is the number of times $c$ occurs in any tweet $t' \in T$. The first factor of this equation reflects the term frequency $TF$, whereas the second factor reflects the inverse document frequency $IDF$ \cite{zangerle2011recommending}.

Based on these similarity values, we identify the most similar tweets $S_t$ for $t$ and extract the hashtags used in these tweets (i.e., $HT_{S_t}$). For each hashtag $ht \in HT_{S_t}$, we assign a content-based score $CB(ht, t)$, which is the highest similarity value within the most similar tweets $S_t$ in which $ht$ occurs. We implement this method using the Lucene-based full-text search engine Apache Solr 4.7.10\footnote{\url{http://lucene.apache.org/solr/}}. Based on Solr's software documentation and our own experimentation, we set the minimum term frequency $tf$ to 2 and the minimum document frequency $df$ to 5.

\para{Combining personalized and content-aware hashtag rec.} We combine our personalized BLL$_{I,S}${} approach with this content-based analysis (C) in order to generate personalized hashtag recommendations (see Figure \ref{fig:approach}). Again, we achieve this via a linear combination of both approaches. Taken together, the top-$k$ recommended hashtags $\widetilde{HT}_{u,t}$ for user $u$ and tweet $t$ are given by:
\begin{equation}
\begin{split}
	\widetilde{HT}_{u,t} = \argmax_{ht \in \overline{HT}_{u,t}}^{k}(\lambda \underbrace{B_{I,S}(ht, u)}_{BLL_{I,S}} + (1 - \lambda) \underbrace{\sigma(CB(ht, t))}_{C})
\end{split}
\end{equation}
where $\overline{HT}_{u,t}$ is the set of candidate hashtags for $u$ and $t$ (i.e., $HT_u \cup HT_{F_u} \cup HT_{S_t}$). The $\lambda$ parameter is used to give weights to the personalized and content-aware components. To that end, we set $\lambda$ to .3 based on experimentation. Please note that the content-based score $CB(ht, t)$ has to be normalized using the softmax function (see Equation \ref{eq:sm}), whereas $B_{I,S}(ht, u)$ is already normalized (see Equation \ref{eq:hybrid}). This finally constitutes our personalized hashtag recommendation algorithm termed BLL$_{I,S,C}${}.\section{Evaluation} \label{sec:evaluation}
In this section, we present the evaluation of our approach. This includes the methodology used as well as the results in terms of recommendation accuracy and ranking for our two scenarios.

\subsection{Methodology}
The methodology of our evaluation is given by the evaluation protocol, evaluation metrics and baseline algorithms used.

\para{Evaluation protocol.} In order to split our datasets into training and test sets, we use an established leave-one-out evaluation protocol from research on information retrieval and recommender systems \cite{jaschke2008tag}. For each seed user in our datasets (see Section \ref{sec:datasets}) with at least two tweets (i.e., 2,020 users in the \textit{CompSci}{} dataset and 2,679 users in the \textit{Random}{} dataset), we determine her most recent tweet and put it (and its hashtags) into the test set. The remaining tweets are then put into the training set. This protocol ensures not only that the hashtags of at least one tweet per user are available for training but also that the chronological order of the data is preserved (i.e., future hashtags are predicted based on usage patterns of past ones). We use these sets in two evaluation scenarios:

\subpara{\textit{Scenario 1}{}.} In the first scenario, we ignore the content of the currently proposed tweet (i.e., the one in the test set) and solely provide hashtag predictions based on the current user-id. Thus, in \textit{Scenario 1}{}, we are able to evaluate all test set tweets.

\subpara{\textit{Scenario 2}{}.} In the second scenario, we also incorporate the content of the current tweet. In this setting, we only evaluate the test set entries, which do not include retweets (i.e., 954 test set tweets in the \textit{CompSci}{} dataset and 1,504 test set tweets in the \textit{Random}{} dataset). The reason for excluding the retweets from the test set in \textit{Scenario 2}{} is that searching for similar tweets in the training set would result in identical tweets with identical hashtags, which would heavily bias our evaluation (see also \cite{zangerle2011recommending}).

\para{Evaluation metrics.} To finally quantify the quality of the algorithms, for each test set entry, we compare the top-$10$ hashtags an algorithm predicts for the given user $u$ and tweet $t$ (i.e., $\widetilde{HT}_{u,t}$) with the set of relevant hashtags actually used by $u$ in $t$.

This comparison is done using various evaluation metrics known from the field of recommender systems. Specifically, we report Precision (P) and Recall (R) for $k$ = 1 to 10 predicted hashtags by means of Precision/Recall plots, and F1-score (F1@5) for $k$ = 5 predicted hashtags. We set $k$ = 5 for the F1-score since F1@5 was also used as the main evaluation metric in the well-known ECML PKDD 2009 discovery challenge\footnote{\url{http://www.kde.cs.uni-kassel.de/ws/dc09/evaluation}.}. Additionally, we report the ranking-dependent metrics Mean Reciprocal Rank (MRR@10), Mean Average Precision (MAP@10) and Normalized Discounted Cumulative Gain (nDCG@10) for $k$ = 10 predicted hashtags \cite{jarvelinMetrics}.

\para{Baseline algorithms.} We compare our approach to a rich set of 9 state-of-the-art hashtag recommendation algorithms:

\begin{table*}[t!]
	\small
  \setlength{\tabcolsep}{6.0pt}	
  \centering
    \begin{tabular}{l||l|lll|lll|llll|lll}
    \specialrule{.2em}{.1em}{.1em}
																		& 					& \multicolumn{10}{c|}{\textit{Scenario 1}{}:}								& \multicolumn{3}{c}{\textit{Scenario 2}{}:}   \\
																		& 					& \multicolumn{10}{c|}{Hashtag rec. w/o current tweet}								& \multicolumn{3}{c}{Hashtag rec. w/ current tweet}   \\
											Dataset				& Metric		& MP$_I$	& MR$_I$	& BLL$_I$				& MP$_S$	& MR$_S$	& BLL$_S$					& MP			& FR		& CF		& BLL$_{I,S}${}		& SR		& TCI		& BLL$_{I,S,C}${}	\\\hline 
											\multirow{4}{*}{\centering{\textit{CompSci}{}}}			                             					              	                                            
																		&	F1@5			& .086		& .098		& \textbf{.101}	& .022		& .076		& \textbf{.118}		& .006		& .083	& .099	& \textbf{.153$^{***}$}		& .139	&	.182	& \textbf{.200$^{*}$}	\\
																		& MRR@10		& .136		& .188		& \textbf{.193}	& .032		& .122		& \textbf{.187}		& .007		& .130	& .163	& \textbf{.268$^{***}$}		& .264	&	.334	& \textbf{.395$^{***}$}	\\
																		& MAP@10		& .143		& .195		& \textbf{.202}	& .033		& .128		& \textbf{.205}		& .007		& .136	& .169	& \textbf{.285$^{***}$}		& .283	&	.354	& \textbf{.417$^{***}$}	\\
																		& nDCG@10		&	.175		& .218		&	\textbf{.225}	&	.046		& .154		&	\textbf{.235}		& .012		&	.169	& .196	&	\textbf{.324$^{***}$}		&	.299	&	.385	&	\textbf{.446$^{**}$} \\\hline																		
											\multirow{4}{*}{\centering{\textit{Random}{}}}																								    		              														          
																		&	F1@5			& .160		& .169		& \textbf{.175}	& .072		& .103		& \textbf{.138}		& .012		& .159	& .165	& \textbf{.208$^{***}$}		& .181	&	.243	& \textbf{.261$^{*}$}	\\
																		& MRR@10		& .261		& .300		& \textbf{.314}	& .109		& .159		& \textbf{.220}		& .023		& .260	& .278	& \textbf{.361$^{***}$}		& .341	&	.436	& \textbf{.489$^{**}$}	\\
																		& MAP@10		& .279		& .315		& \textbf{.335}	& .116		& .171		& \textbf{.240}		& .024		& .279	& .296	& \textbf{.389$^{***}$}		& .374	&	.472	& \textbf{.530$^{**}$}	\\
																		& nDCG@10		&	.323		& .352		&	\textbf{.370}	&	.144		& .205		&	\textbf{.280}		& .035		&	.324	& .333	&	\textbf{.434$^{***}$}		&	.388	&	.507	&	\textbf{.562$^{**}$}	\\
		\specialrule{.2em}{.1em}{.1em}								
    \end{tabular}
    \caption{Recommender accuracy results of our two evaluation scenarios. In \textit{Scenario 1}{}, we compare approaches that ignore the current tweet content, while in \textit{Scenario 2}{}, we compare algorithms that also incorporate the current tweet. We observe that (i) BLL$_I$ outperforms MP$_I$ and MR$_I$, (ii) BLL$_S$ outperforms MP$_S$ and MR$_S$, (iii) BLL$_{I,S}$ outperforms MP, FR and CF, and (iv) BLL$_{I,S,C}${} outperforms SR and TCI. Based on a t-test, the symbols $^{*}$ ($\alpha$ = .1), $^{**}$ ($\alpha$ = .01) and $^{***}$ ($\alpha$ = .001) indicate statistically significant differences between BLL$_{I,S}${} and CF in \textit{Scenario 1}{}, and between BLL$_{I,S,C}${} and TCI in \textit{Scenario 2}{}.
\vspace{-3mm}}	
  \label{tab:results}
\end{table*}

\subpara{MP$_I$.} The \textit{Most Popular Individual Hashtags} algorithm ranks the hashtags based on the frequency in the hashtag assignments of current user $u$. MP$_I$ is also referred to as Most Popular Tags by User (MP$_u$) in tag recommendation literature \cite{jaschke2008tag}.

\subpara{MR$_I$.} \textit{Most Recent Individual Hashtags} is a time-dependent variant of MP$_I$. MR$_I$ suggests the $k$ most recently used hashtags of current user $u$ \cite{campos2014time}. Our BLL$_I$ approach can be seen as an integrated combination of MP$_I$ and MR$_I$ based on human memory theory.

\subpara{MP$_S$.} The \textit{Most Popular Social Hashtags} algorithm is the social correspondent to the individual MP$_I$ approach \cite{jaschke2008tag}. Thus, MP$_S$ does not rank the hashtags based on the frequency in the hashtag assignments of user $u$ but based on the frequency in the hashtag assignments of user $u$'s set of followees $F_u$.

\subpara{MR$_S$.} \textit{Most Recent Social Hashtags} is the time-dependent equivalent to MP$_S$. MR$_S$ sorts the hashtag assignments of $u$'s followees $F_u$ by time and recommends the $k$ most recent ones. Our BLL$_S$ algorithm is a cognitive-inspired integration of MP$_S$ and MR$_S$.

\subpara{MP.} The unpersonalized \textit{Most Popular Hashtags} approach returns the same set of hashtags for any user. These hashtags are ranked by their overall frequency in the dataset \cite{jaschke2008tag}.

\subpara{FR.} \textit{FolkRank} is an adaption of Google's PageRank approach used to rank the entities in folksonomy graphs and has become one of the most successful tag recommender methods \cite{hotho2006folkrank}. %
We use the standard FR implementation provided by the University of Kassel\footnote{\url{http://www.kde.cs.uni-kassel.de/code}} with its suggested default parameters. More specifically, the weight of the preference vector $d$ is set to .7 and the maximum number of iterations $l$ is set to 10 \cite{jaschke2008tag}.

\subpara{CF.} \textit{User-based Collaborative Filtering} is a well-known algorithm used in many variants of modern recommender systems and was adapted by \cite{marinho2008collaborative} for use in tag-based settings. We apply the same idea for the task of recommending hashtags and thus, first identify the $k$ most similar users (i.e., the nearest neighbors) for current user $u$ by means of the cosine similarity measure and then suggest the hashtags used by these neighbors. For our experiments, we use a neighborhood size $k$ of 20 users (see also \cite{gemmell2009improving}).

\subpara{SR.} \textit{SimilarityRank} is an unpersonalized hashtag recommendation algorithm, which utilizes the content of the currently proposed tweet $t$ \cite{zangerle2011recommending}. Similarly to our BLL$_{I,S,C}${} approach, this is achieved using TF-IDF to determine content-based similarity scores between tweets (see Section \ref{sec:hashtagrec}). These scores are used to recommend the $k$ hashtags that occur in $t$'s most similar tweets.

\subpara{TCI.} \textit{TemporalCombInt} is one of the most recent approaches for personalized hashtag recommendations and also one of the very few approaches that accounts for the effect of time on hashtag usage \cite{harvey2015long} (see also Section \ref{sec:relatedwork}). TCI builds on a linear combination of SR and CF and incorporates temporal effects by considering the time-dependent relevance of a hashtag with respect to the recommendation date. This is done by categorizing the hashtags into ``organizational'' and ``conversational'' hashtags, and modeling the decay of temporal relevance using an exponential function. By fitting this model to our crawled data, we set the two main parameters of the algorithm, $\eta_l$ and $\eta_h$, to .1 and .2, respectively. 

\subsection{Results and Discussion} \label{sec:results}
\begin{figure*}[t!]
   \centering
	 \captionsetup[subfigure]{justification=centering}
	 \subfloat[][\textit{Scenario 1}{}: Hashtag rec. w/o current tweet\\\textit{CompSci}{} dataset]{ 
      \includegraphics[width=0.24\textwidth]{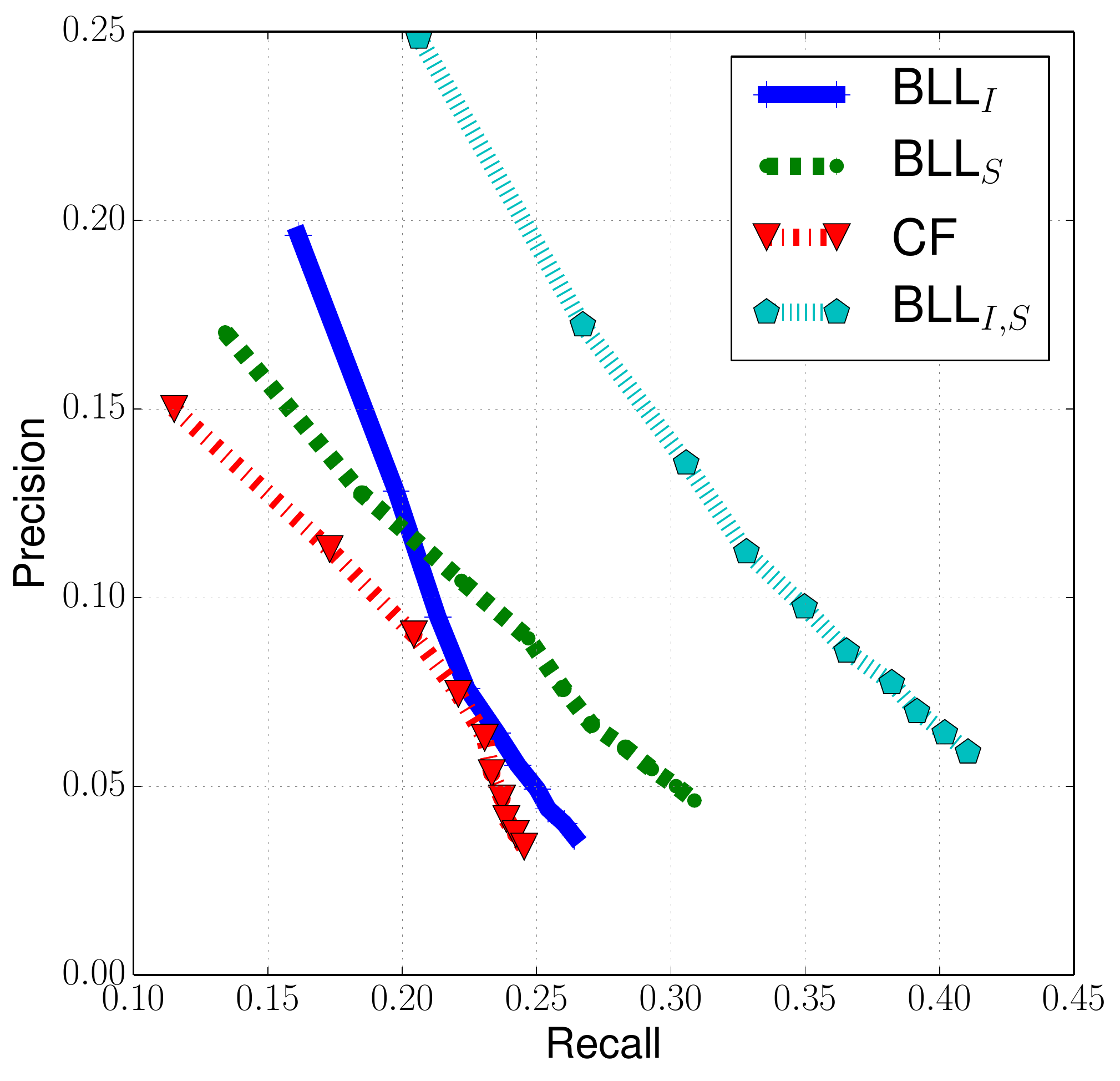} 
   }
	 \subfloat[][\textit{Scenario 1}{}: Hashtag rec. w/o current tweet\\\textit{Random}{} dataset]{ 
      \includegraphics[width=0.24\textwidth]{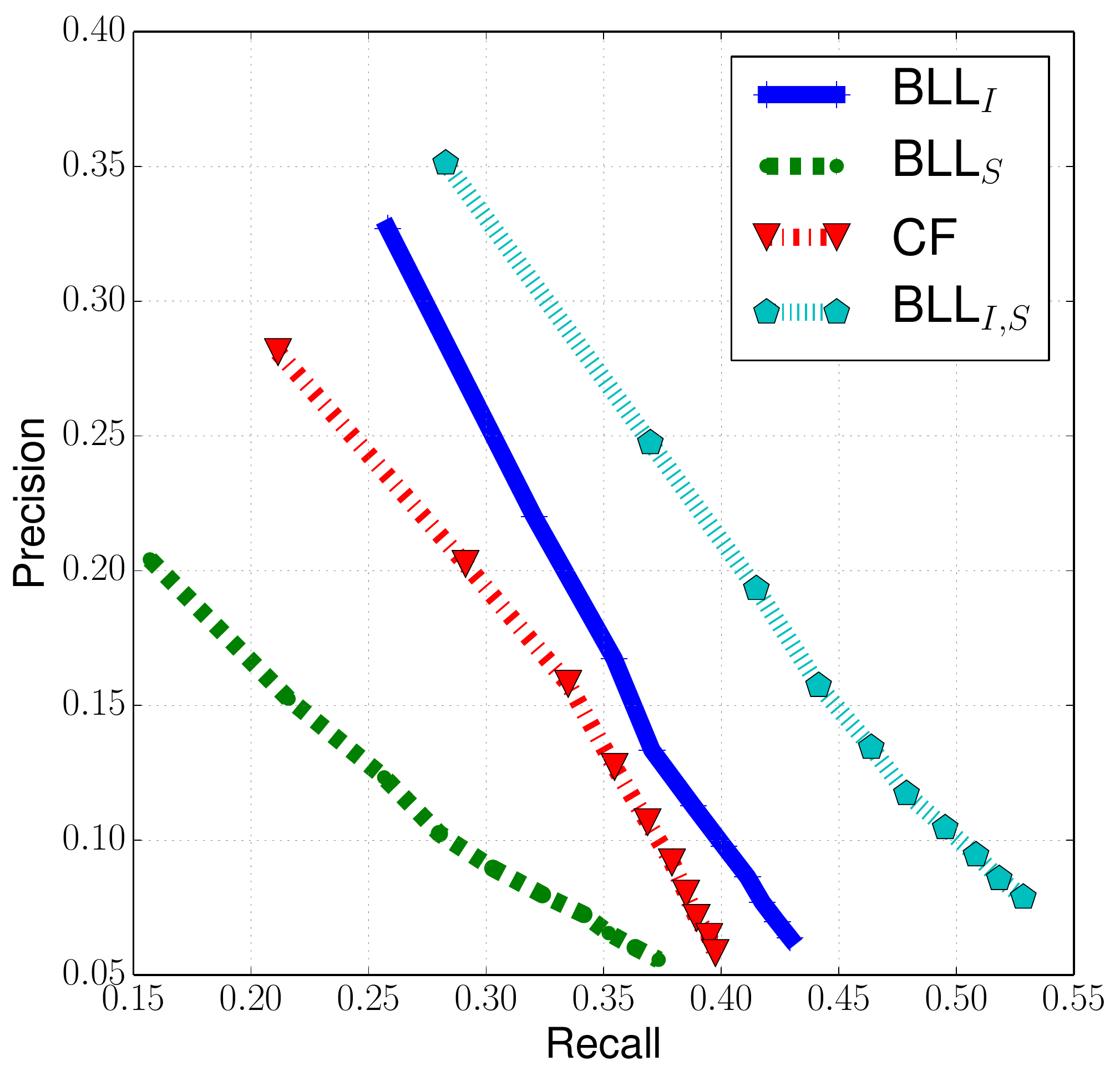} 
   }
	 \subfloat[][\textit{Scenario 2}{}: Hashtag rec. w/ current tweet\\\textit{CompSci}{} dataset]{ 
      \includegraphics[width=0.24\textwidth]{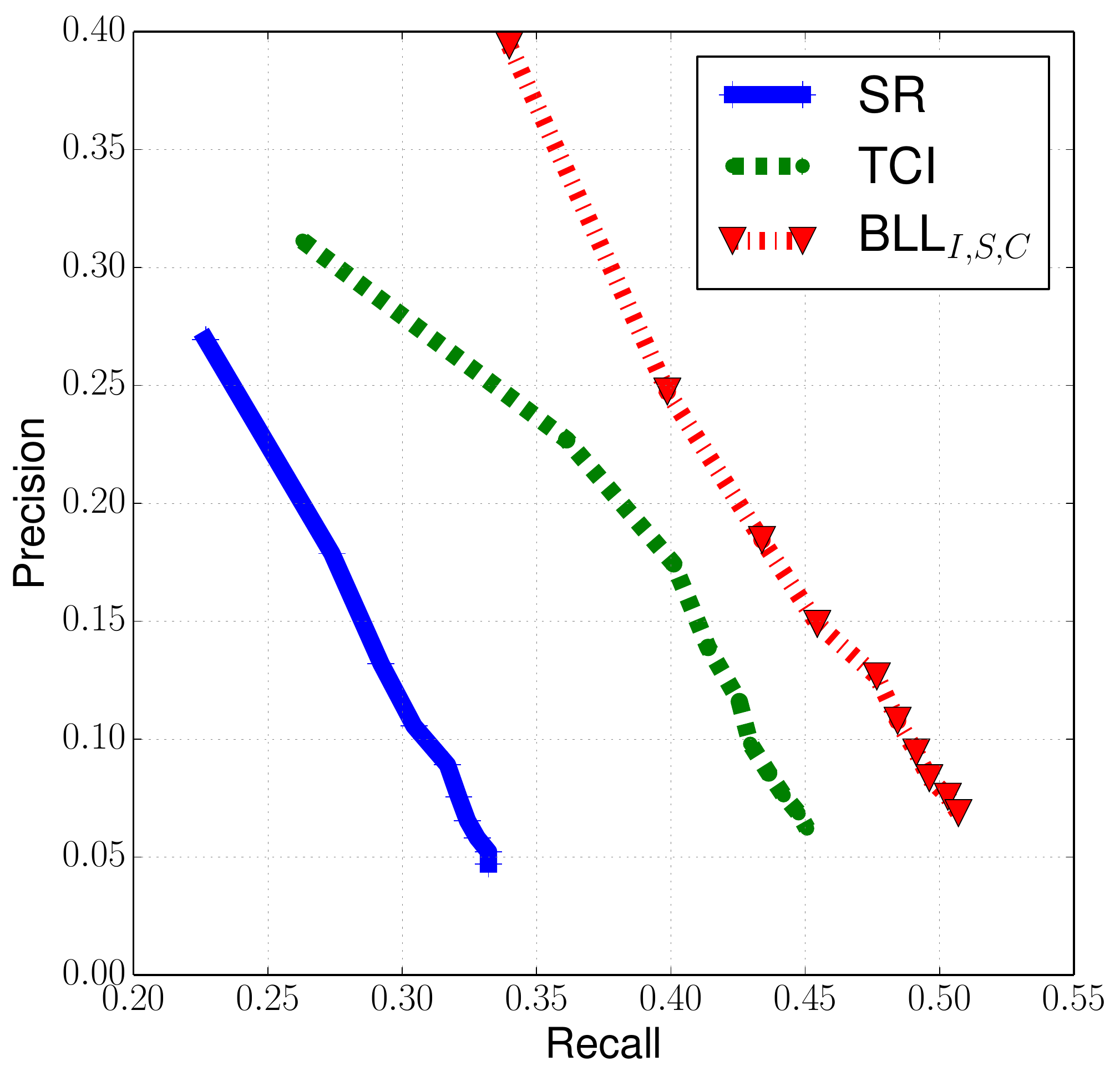} 
   }
	 \subfloat[][\textit{Scenario 2}{}: Hashtag rec. w/ current tweet\\\textit{Random}{} dataset]{ 
      \includegraphics[width=0.24\textwidth]{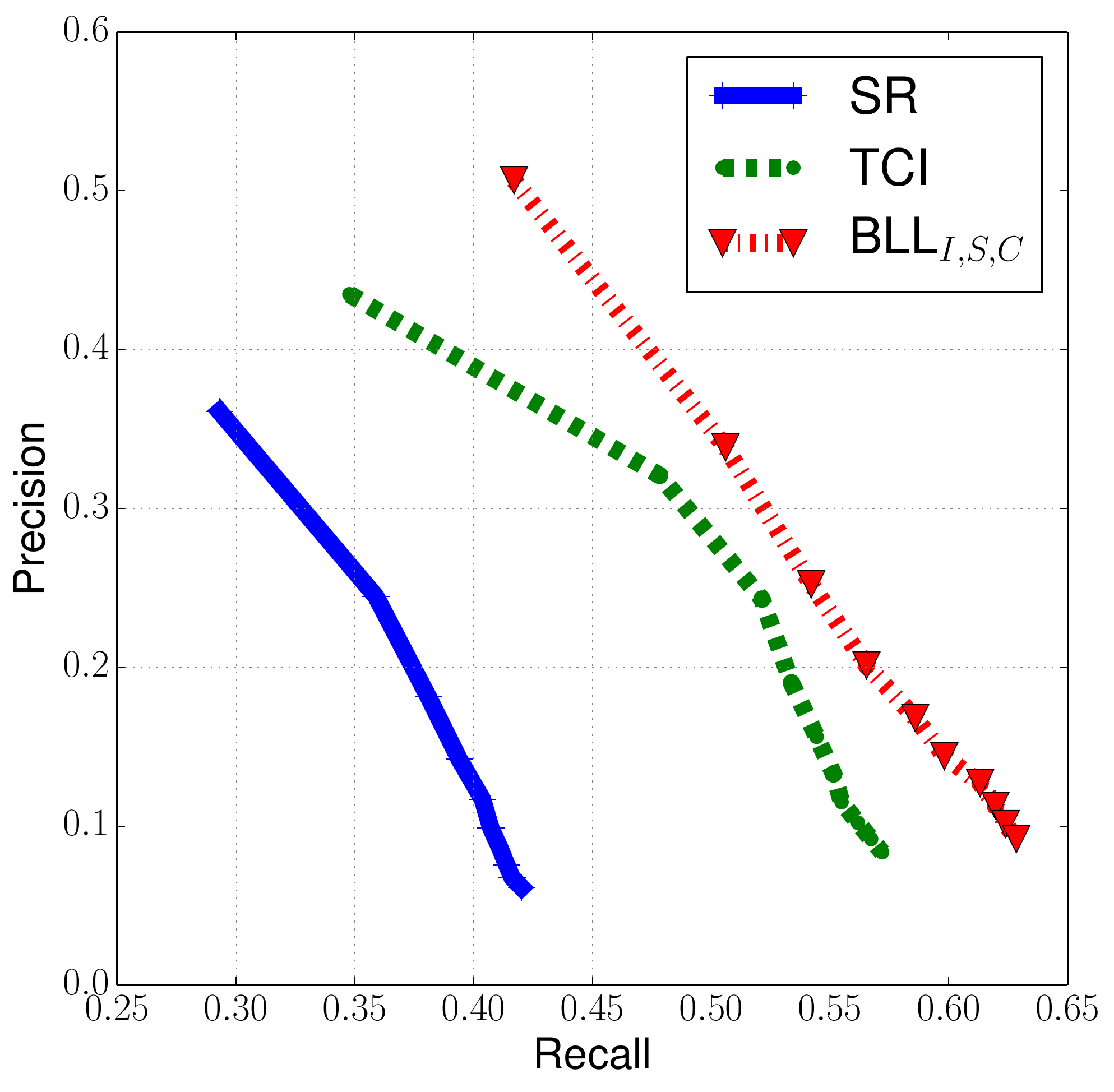} 
   }
   \caption{Precision / Recall plots of our two evaluation scenarios showing the accuracy of BLL$_I$, BLL$_S$, CF, BLL$_{I,S}${}, SR, TCI and BLL$_{I,S,C}${} for $k$ = 1 - 10 recommended hashtags. Again, BLL$_{I,S}${} provides the best results in \textit{Scenario 1}{} and BLL$_{I,S,C}${} in \textit{Scenario 2}{}.
\vspace{-3mm}}
	 \label{fig:results}
\end{figure*}

In Section \ref{sec:analysis}, we found that time is an important factor for hashtag reuse. Because of this, we assume that our time-dependent and cognitive-inspired approach should provide reasonable results compared to other algorithms. The accuracy estimates for our two evaluation scenarios are shown in Table \ref{tab:results} and Figure \ref{fig:results}.

\para{\textit{Scenario 1}{}: Hashtag rec. w/o current tweet.} In our first evaluation scenario, we validate approaches that predict future hashtags without incorporating the content of the currently proposed tweet. Here, we identify three main results:

\subpara{(a) BLL$_I$ $>$ MP$_I$, MR$_I$.} When predicting individual hashtag reuse, we compare our BLL$_I$ approach to the frequency-based MP$_I$ and the recency-based MR$_I$ algorithms. The results clearly reflect the importance of the time component since MR$_I$ and BLL$_I$ provide higher prediction accuracy and ranking estimates than MP$_I$ for all evaluation metrics across both datasets. Apart from that, we observe that BLL$_I$ outperforms MR$_I$, which speaks in favor of the cognitive-inspired combination of hashtag frequency and recency by means of the BLL equation.

\subpara{(b) BLL$_S$ $>$ MP$_S$, MR$_S$.} Concerning the prediction of social hashtag reuse, we compare our BLL$_S$ approach to the frequency-based MP$_S$ and the recency-based MR$_S$ methods. Similar to the case of individual hashtag reuse, MR$_S$ and our BLL-based method provide higher accuracy estimates than the solely frequency-based one, but interestingly, this time the differences between these methods is much larger. This indicates that the time information is especially important in a social setting. We somehow expected this behavior since typically only the most recent tweets of the followees are shown on a user's Twitter timeline. Again, the combination of hashtag frequency and recency by means of the BLL equation provides the best results.

\subpara{(c) BLL$_{I,S}${} $>$ MP, FR, CF.} Finally, we compare our hybrid BLL$_{I,S}${} approach to the unpersonalized MP algorithm, the well-known FR method from tag recommender research and classic user-based CF. The first observation that becomes apparent is the poor performance of the unpersonalized MP baseline, which underpins the importance of personalized methods for hashtag recommendation.
Additionally, and more importantly, our hybrid BLL$_{I,S}${} approach does not only improve its BLL$_I$ and BLL$_S$ components but also provides significantly higher accuracy and ranking estimates than FR and CF. This shows that BLL$_{I,S}${} is capable of providing reasonable hashtag recommendations solely based on temporal usage patterns of past hashtag assignments.

\para{\textit{Scenario 2}{}: Hashtag rec. w/ current tweet.} In the second scenario, we evaluate hashtag recommendation methods that also incorporate the content of the current tweet. This includes the unpersonalized SR approach, the time-dependent TCI algorithm and our BLL$_{I,S,C}${} approach. Our two main results are:

\subpara{(a) TCI, BLL$_{I,S,C}${} $>$ SR.} The first main result of our second evaluation scenario is that both time-dependent methods TCI and BLL$_{I,S,C}${} outperform the unpersonalized SR approach. We somehow expected this result since both TCI and BLL$_{I,S,C}${} extend the TF-IDF-based tweet content analysis of SR with personalization techniques via CF (TCI) or the BLL equation (BLL$_{I,S,C}${}).

\subpara{(b) BLL$_{I,S,C}${} $>$ TCI.} The second main result of \textit{Scenario 2}{} is that BLL$_{I,S,C}${} provides significantly higher accuracy estimates than TCI. This is due to three main differences between these methods: (i) instead of using hashtags of similar users by means of CF for adding personalization, we incorporate not only individual hashtags of the current user but also social hashtags of the current user's followees, (ii) instead of applying the effect of time on a global hashtag level, we model the time-dependent decay on an individual and social level, and (iii) instead of modeling this time-dependent decay using an exponential function, we use a power function by means of the BLL equation.

\para{\textit{CompSci}{} dataset vs. \textit{Random}{} dataset.} Another interesting finding we observe is that all algorithms provide better results for the \textit{Random}{} dataset than for the \textit{CompSci}{} dataset. In our case, this indicates that the task of predicting hashtags in the domain-specific network of computer scientists is harder than in the network of random users. If we look back at Figure \ref{fig:intro}, this makes sense since the amount of ``external'' hashtags is twice as high in the \textit{CompSci}{} dataset (i.e., 26\%) than in the \textit{Random}{} one (i.e., 13\%).

\vspace{-1mm}
\findingbox{The BLL equation, which accounts for temporal effects of item exposure in human memory, provides a suitable model for personalized hashtag recommendations. This is validated in two evaluation scenarios (i.e., without and with incorporating the content of the current tweet), in which our cognitive-inspired approach outperforms several state-of-the-art hashtag recommendation algorithms in terms of prediction accuracy.}

\section{Related Work} \label{sec:relatedwork}
Over the past years, tagging has emerged as an important feature of the social Web, which supports users to collaboratively organize and find content \cite{Korner2010}. Two types of tags have been established: (i) social tags as used in systems like BibSonomy and CiteUlike, and (ii) hashtags as used in systems like Twitter and Instagram. Whereas social tags are mainly used to index resources for later retrieval, hashtags have a more conversational nature and are used to filter and direct content to certain streams of information \cite{Huang2010}.

One of the most prominent approaches in the field of tag recommendations is the FolkRank algorithm \cite{hotho2006folkrank,jaschke2007tag,jaschke2008tag}. FolkRank is an extension of the well-known Google PageRank approach to rank the entities in folksonomies (i.e., users, resources and tags). Other important tag recommendation methods are based on Collaborative Filtering \cite{marinho2008collaborative,gemmell2009improving}, Latent Dirichlet Allocation \cite{krestel2009latent,krestel2012personalized} or Tensor Factorization \cite{rendle2010pairwise,rendle2009learning}. Recent observations in the field of social tagging state the importance of the time component for the individual tagging behavior of users. In this respect, \cite{zhang2012integrating,yin2011exploiting,yin2011temporal} propose time-dependent tag recommender approaches, which model the tagging variation over time using exponential functions. In our previous work \cite{www_bll,Kowald2016a}, we presented a more theory-driven approach, where we use the BLL equation coming from the cognitive architecture ACT-R \cite{anderson_reflections_1991,anderson2004integrated} to model the power-law of time-dependent decay. We evaluated our approach in detail and compared it to other state-of-the-art methods in \cite{Kowald2015}. In the present work, we build upon our results and incorporate the BLL equation to study the effect of time on hashtag reuse to design our hashtag recommendation approach.

There is already a large body of research available that focuses on the recommendation of hashtags in Twitter. One illustrative example is the work presented in \cite{Godin2013}, in which hashtag recommendations are provided by categorizing tweets into general topics using LDA. The approach then recommends the hashtags that best fit the topics of a new tweet. The authors evaluate their approach using a qualitative study, in which they ask persons if the recommended hashtags describe the topics of a tweet and could be used to semantically enrich it. In 80\% of the cases, they are able to provide a suitable hashtag from a selection of five possibilities. Other similar approaches that use topic models for hashtag recommendations are presented in \cite{She2014,wang2014tag,xu2015personalized,efron2010hashtag}. In \cite{jeon2014hashtag}, a related algorithm based on a hashtag classification scheme is proposed. 
The most notable work in the context of hashtag recommendations is probably the content-based SR approach presented in \cite{zangerle2011recommending} and \cite{zangerle2013impact}. The authors use the TF-IDF statistic to calculate similarities between tweets and identify suitable hashtags based on these similarity scores. They show that SR improves Recall and Precision by around 35\% compared to a popularity-based approach. Our BLL$_{I,S,C}${} approach uses the same statistic to integrate the content of a user's currently proposed tweet. In \cite{kywe2012recommending}, a personalized extension of SR is presented, in which the authors combine it with user-based CF. Apart from that, a content-based hashtag recommendation algorithm for hyper-linked tweets is proposed in \cite{sedhai2014hashtag}.

Related research has studied temporal effects on hashtag usage, for instance in the context of popular hashtags in Twitter \cite{lin2012study,lehmann2012dynamical,tsur2012,ma2012will}. For example, in \cite{ma2012will}, the authors aim to predict if a specific hashtag will be popular on the next day. By formulating this task as a classification problem, they find that both content features (e.g., the topic of the hashtag) and context features (e.g., the users who used the hashtags) are effective features for popularity prediction. A similar approach is presented in \cite{yang2011patterns}, in which the authors uncover the temporal dynamics of online content (e.g., tweets) by formulating a time series clustering problem. One of the very few examples of a time-aware hashtag recommendation approach is the recently proposed algorithm described in \cite{harvey2015long}. The authors extend the content-based SR approach \cite{zangerle2011recommending} with a personalization technique by means of CF and further consider the temporal relevance of hashtags. To account for this temporal relevance, they divide the hashtags into two categories: ``organizational'' ones, which are used over a long period of time and ``conversational'' ones, which are used only during a short time span (e.g., for a specific event).

In contrast to our proposed algorithm, which relies on the BLL equation, their approach considers the effect of time on a global hashtag level of the whole Twitter network and not on an individual and social level of a specific user. Furthermore, we use a power function rather than an exponential one to model the time-dependent decay based on our empirical findings.\section{Conclusion and Future Work} \label{sec:conclusion}
In this paper, we presented a cognitive-inspired approach for hashtag recommendations in Twitter. Our approach utilizes the BLL equation from the cognitive architecture ACT-R to account for temporal effects on individual hashtag reuse (i.e., reusing own hashtags) and social hashtag reuse (i.e., reusing hashtags, which has been previously used by a followee). Our analysis of hashtag usage types in two empirical networks (i.e., \textit{CompSci}{} and \textit{Random}{} datasets) crawled from Twitter reveals that between 66\% and 81\% of hashtag assignments can be explained by past individual and social hashtag usage. By analyzing the timestamps of these hashtag assignments, we find that temporal effects play an important role for both individual and social reuse of hashtags and that a power function provides a better fit to model this time-dependent decay than an exponential function.

Thus, the more recently a hashtag was used by a user or her followees, the higher the probability that this user will use the same hashtag again later in time. Based on these findings, we utilized the Base-Level Learning (BLL) equation of the cognitive architecture ACT-R, which accounts for the time-dependent decay of item exposure in human memory, to develop BLL$_{I,S}${} and BLL$_{I,S,C}${}, two algorithms for recommending hashtags. Whereas BLL$_{I,S}${} aims to recommend hashtags without incorporating the current tweet (\textit{Scenario 1}{}), BLL$_{I,S,C}${} also utilizes the content of the current tweet using the TF-IDF statistic (\textit{Scenario 2}{}). We compared both algorithms to state-of-the-art hashtag recommendation algorithms and found that our cognitive-inspired approaches outperform these algorithms in terms of prediction accuracy and ranking.

One limitation of this work is that we model the reuse of social hashtags solely by analyzing how frequently and recently a hashtag was used by a user's followees, neglecting by whom the hashtag was used. Thus, for future work, we plan to extend our approach with the social status of the followee (e.g., via the reputation of the user by means of the number of followers). In this respect, we will also utilize the social connection strength between a user and her followee (e.g., by the number of mentions or retweets).

With respect to the hashtag assignments that cannot be explained by hashtag reuse (i.e., 26\% in the \textit{CompSci}{} dataset and 13\% in the \textit{Random}{} dataset), we want to utilize an external knowledge base to also account for these hashtag assignments. We will achieve this by suggesting hashtags of currently trending topics or events. Finally, we also plan to verify our findings in larger Twitter data samples than the ones used in this paper as well as in other online social networks that feature hashtags, such as Instagram and Facebook.

In summary, our work contributes to the rich line of research on improving the use of hashtags in social networks. We hope that future work will be attracted by our insights into how temporal effects on hashtag usage can be modeled using models from human memory theory, such as the BLL equation.

\para{Acknowledgments.} The authors would like to thank Matthias Traub and Dieter Theiler for valuable inputs. This work is funded by the Know-Center and the EU project AFEL (GA: 687916).

\small
\bibliographystyle{abbrv}

\end{document}